\documentclass[sigconf]{acmart}

\usepackage{placeins}
\usepackage{dblfloatfix} 
\usepackage{amsmath}
\usepackage{amsthm}
\usepackage{graphicx}
\usepackage{hyperref}
\usepackage{booktabs}
\usepackage{xcolor}
\usepackage[most]{tcolorbox}
\usepackage{bm}
\usepackage{algorithmic}
\usepackage{textcomp}

\newtheorem{definition}{Definition}

\DeclareMathOperator*{\argmax}{argmax}

\AtBeginDocument{%
  }

\setcopyright{none}
\copyrightyear{2025}
\acmYear{2025}
\acmDOI{XXXXXXX.XXXXXXX}

\acmConference[MICRO SPICE '25]{Workshop}{October 2025}{Seoul, Korea}
\acmISBN{978-X-XXXX-XXXX-X/XX/XX}

\tcbset{
  mybluebox/.style={
    colback=blue!3!white,
    colframe=blue!60!white,
    coltitle=blue!15!white,
    fonttitle=\bfseries,
    boxrule=0.8pt,
    arc=2pt,
    left=6pt,right=6pt,top=6pt,bottom=6pt,
    title style={left
    color=blue!10!white,right color=blue!5!white},
    before skip=8pt plus 2pt,
    after skip=8pt plus 2pt,
    breakable,
    enhanced jigsaw
  }
}

\newtcolorbox{assumpbox}[1][]{mybluebox,title={Assumptions for Claim A},#1}
\newtcolorbox{lemmabox}[1][]{mybluebox,title={Lemma A},#1}
\newtcolorbox{theorembox}[1][]{mybluebox,title={Theorem A (Claim A—Existence \& Irregularity)},#1}
\newtcolorbox{corobox}[1][]{mybluebox,title={Corollary A},#1}

\newtcolorbox{assumpboxC}[1][]{mybluebox,title={Assumptions for Claim C},#1}
\newtcolorbox{lemmaboxC}[1][]{mybluebox,title={Lemma C},#1}
\newtcolorbox{theoremboxC}[1][]{mybluebox,title={Theorem C},#1}
\newtcolorbox{coroboxC}[1][]{mybluebox,title={Corollary C},#1}
\begin{document}

\title{Taming the Tail: NoI Topology Synthesis for Mixed DL Workloads on Chiplet-Based Accelerators}

\author{Arnav Shukla}
\affiliation{%
  \institution{Indraprastha Institute of Information Technology Delhi}
  \state{New Delhi}
  \country{India}
}
\email{arnav22103@iiitd.ac.in}

\author{Harsh Sharma}
\affiliation{%
  \institution{Washington State University}
  \city{Pullman}
  \state{Washington}
  \country{USA}
}
\email{harsh.sharma@wsu.edu}

\author{Srikant Bharadwaj}
\affiliation{%
  \institution{Microsoft Research}
  \city{Redmond}
  \state{Washington}
  \country{USA}
}
\email{srikant.bharadwaj@microsoft.com}

\author{Vinayak Abrol}
\affiliation{%
  \institution{Indraprastha Institute of Information Technology Delhi}
  \state{New Delhi}
  \country{India}
}
\email{abrol@iiitd.ac.in}

\author{Sujay Deb}
\affiliation{%
  \institution{Indraprastha Institute of Information Technology Delhi}
  \state{New Delhi}
  \country{India}
}
\email{sdeb@iiitd.ac.in}

\renewcommand{\shortauthors}{Shukla et al.}

\begin{abstract}
Heterogeneous chiplet-based systems improve scaling by disaggregating CPUs/GPUs and emerging technologies (HBM/DRAM). However this on-package disaggregation introduces a latency in Network-on-Interposer(NoI). 
We observe that in modern large-model \emph{inference}, parameters and activations routinely move back and forth from HBM/DRAM, injecting large, bursty flows into the interposer. 
These memory-driven transfers inflate tail latency and violate Service Level Agreements (SLAs) across k-ary n-cube baseline NoI topologies. To address this gap we introduce an Interference Score (IS) that quantifies worst-case slowdown under contention. We then formulate NoI synthesis as a multi-objective optimization (MOO) problem. We develop PARL (Partition-Aware Reinforcement Learner), a topology generator that balances throughput, latency, and power. PARL-generated topologies reduce contention at the memory cut, meet SLAs, and cut worst-case slowdown to $1.2\times$ while maintaining competitive mean throughput relative to link-rich meshes. Overall, this reframes NoI design for heterogeneous chiplet accelerators with workload-aware objectives.

\end{abstract}

\maketitle
\section{Introduction}
Chiplet-based design circumvents reticle and cost limits by pairing advanced-node compute with mature-node I/O/memory, as in AMD EPYC/Instinct \cite{b1,b2,b3,b4}. Disaggregation elevates the Network-on-Interposer (NoI) to a first-order bottleneck \cite{b5}.

Modern DL—especially Mixture-of-Experts (MoE)—stresses the NoI not only by volume but by \emph{structure}. Models exceed on-chiplet SRAM and often on-package HBM; parameters/
activations spill to shared HBM or DRAM, injecting heavy-tailed, bursty, often multicast traffic \cite{b6,b7,b8}. In MoE, conditional 
compute activates $k \ll E$ experts per token \cite{b19,b20}—e.g., Mixtral-8x7B decouples its 47B total parameters from per-token FLOPs while demanding 256 MB per expert from memory. Large production deployments (e.g., Mixtral-style top-$k$ gating) push footprints beyond local caches \cite{b21,b22}, yielding:
\begin{enumerate}
  \item \textbf{Burst-amplified memory traffic:} gating in MoE drives variance, causing short surges at memory-adjacent routers.
  \item \textbf{Spatially fragmented parameters:} experts spread across chiplets force frequent HBM fetches across the memory cut.
\end{enumerate}
Unlike dense transformers dominated by regular collectives, MoE mixes sparse many-to-few gathers, few-to-many scatters, and weight streaming. Diameter/bisection proxies miss these localized queues, motivating a tail-aware objective and automated synthesis.

\textbf{Contributions.}
\begin{itemize}
  \item \textbf{Tail-aware formulation \& metric:} we characterize mixed, memory-sourced traffic and introduce a novel \emph{Interference Score (IS)} which models worst-case degradation.
  \item \textbf{PARL topology generator:} a partition-aware RL framework that synthesizes NoI under a multi-objective reward.
  \item \textbf{Evaluation:} PARL sustains SLAs and improves robustness over k-ary n-cube baselines that exhibit p95/p99 blow-ups.
\end{itemize}
\section{Related Work}
Handcrafted NoI topologies largely follow classical k-ary n-cubes. HexaMesh reduces diameter raising bisection bandwidth versus grids \cite{b9}; KITE accurately models heterogeneous interposer links \cite{b5}. These frameworks optimize global or mean-case metrics, not tail risk under mixed, memory-sourced traffic. SWAP co-designs compute/memory placement with the interconnect for server-scale accelerators but does not model memory injection \cite{b10}. Recent studies profile fine-grained communication patterns as unicast/multicast mixes on large multi-chiplet AI systems, identifying multicast and memory-driven movement as key scalability bottlenecks \cite{b8}. Classical heavy-traffic queueing theory (e.g., Kingman) predicts superlinear growth near saturation and under high arrival \cite{b12}. We apply this analysis at memory-adjacent routers to explain early p95/p99 blow-ups and to motivate partition-aware, tail-focused synthesis.

\section{Inevitability of Irregular \& Bursty Traffic}
\subsection{Irregular HBM Traffic}
We first prove that irregular, memory-sourced traffic is inevitable under large-scale DL models on chiplet-based hardware.
To this aim, consider the following system model and foundational assumption:\\
\noindent\textbf{Configuration} [MI300X-Class Memory Hierarchy] \label{def:mi300x_memory}
Based on AMD specifications \cite{b11}, each chiplet (XCD) provides: \textbf{L2 cache}: 4 MB per XCD, \textbf{Infinity Cache}: 256 MB shared across 8 XCDs and \textbf{HBM3}: 192 GB total, $\sim$5.3 TB/s peak bandwidth.
\vspace{-10pt}
\begin{definition}[Expert Structure and MoE Routing] \label{def:expert_moe}
Throughout we model use Mixtral-8x7B (32 layers, d=4096, 8 experts with top-2 routing, 32k context). Each MoE expert is a neural network with shape $d \to 4d \to d$ and parameter count $P_{\text{FFN}} = 8d^2$. For a batch of $B$ tokens, a router selects top-$k$ experts per token. Let $I_{t,e} \in \{0,1\}$ indicate if token $t$ routes to expert $e$. The expert load is $N_e = \sum_{t=1}^{B} I_{t,e}$. Under noisy gating, load statistics are $\mathbb{E}[N_e] = B \cdot p_e$ and $\mathrm{Var}[N_e] > 0$ for any expert with selection probability $0 < p_e < 1$. We denote $S_j$ as effective on-die SRAM budget for chiplet $j$.
\end{definition}
\begin{tcolorbox}[colback=blue!10,colframe=blue!50!black,title=\textbf{Claim A: Inevitability of Irregular HBM Traffic}]
\label{claim:a}
For a chiplet platform executing a sparse MoE transformer layer, \textbf{any mapping} $\pi$ sustaining non-zero throughput must have a non-empty subset of compute chiplets where data is sourced from HBM. Thus proving \textit{existence} claim. Since MoE routing induces \textbf{random per-expert token counts}, the HBM-sourced injection rate is \textbf{time-varying and irregular}.
\end{tcolorbox}

\textit{Proof} We break the proof as:

\noindent\textbf{Step 1 - Expert Weights Overflow:} \label{lemma:weight_overflow}

\noindent \textit{The parameter bytes of one expert, $W_{\text{FFN}} = 8d^2 \cdot s$ (where $s \in \{2,4\}$ bytes/weight), exceed the L2 cache of any single XCD and are comparable to the shared Infinity Cache.}

{Proof:} Using $d=4096$ and low-precision storage $s=2$ bytes (BF16), an expert's parameter footprint is $W_{\text{FFN}} = 8d^2s = 256$ MB, which exceeds a single XCD's 4 MB L2 cache and consumes the entire 256 MB shared Infinity Cache. This value exceeds a single XCD's 4\,MB L2 cache and consumes the shared Infinity Cache, so expert weights for any non-trivial mapping must reside in HBM.

\noindent\textbf{Step 2 - Routing Induces Non-Zero Variance:}\label{lemma:routing_variance}

\noindent \textit{Under noisy top-$k$ routing with load-balancing, expert $e$ with selection probability $0 < p_e < 1$, the variance of its load is: $\mathrm{Var}[N_e] > 0$.}

\noindent \textit{Proof:} For independent token-wise routing, $\mathrm{Var}[N_e] = B \cdot p_e(1-p_e) > 0$ whenever $0 < p_e < 1$. Noise and capacity constraints prevent collapse to deterministic routing.

\noindent \textbf{Step 3 - Random Loads Induce Irregular Injection:}

\noindent Let $B_e(\Delta)$ be the HBM-sourced bytes for expert $e$ during time window $\Delta$. This traffic scales with the number of tokens routed to it: $B_e(\Delta) \geq \gamma + \beta \cdot N_e(\Delta)$, where $\gamma$ is a fixed cost and $\beta > 0$ is the per-token data requirement. The variance of this traffic is thus:
$\mathrm{Var}[B_e(\Delta)] \geq \beta^2 \cdot \mathrm{Var}[N_e(\Delta)] > 0$.
Since the injection rate $B_e(\Delta)/|\Delta|$ has non-zero variance, it is not constant. Hence the HBM injection from all model operations is irregular.

\begin{tcolorbox}[colback=green!10, colframe=green]
\textbf{Takeaway 1:} We established that HBM injection is \textbf{inevitable} and variance analysis proves this injection is inherently \textbf{irregular} due to MoE's stochastic routing.
\end{tcolorbox}

\subsection{Catastrophic Interference at the Memory Cut}
We now demonstrate how the irregular traffic identified in Claim A leads to catastrophic interference and tail latency at the memory-to-compute network cut. We use a multi-commodity flow (MCF) and a queueing theory model.
\begin{definition}[Regular NoI Topology] \label{def:regular_noi}
A package/NoI graph $G=(C,E)$ with link capacities $b_e$ is \textit{regular} if each node has degree $d$ and all link capacities are bounded.
\end{definition}

\begin{definition}[Memory Cut] \label{def:memory_cut}
For a set of memory chiplets $M \subset C$, the \textbf{memory cut} is the set of edges connecting memory to non-memory chiplets: $\mathcal{C}_M = \{(u,v) \in E: u \in M, v \notin M\}$. Its capacity is $\text{Cap}(\mathcal{C}_M) = \sum_{e \in \mathcal{C}_M} b_e$.
\end{definition}

\begin{definition}[Arrival and Service Variability] \label{def:variability}
MoE gating produces bursty arrivals with squared coefficient of variation $C_a^2 > 0$. HBM service exhibits variability $C_s^2 > 0$ due to refresh, write-drain, and bank conflicts.
\end{definition}

\begin{tcolorbox}[colback=blue!10,colframe=blue!50!black,title=\textbf{Claim B: Memory Cut Bottleneck and Tail Latency Explosion}]
\label{claim:b}
Consider a regular NoI with memory chiplets $M$. The sustainable MoE throughput $R$ (tokens/s) is bounded by the memory cut capacity. Moreover, queues at memory-adjacent routers experience superlinear growth in tail latency as utilization $\rho \to 1$, governed by Kingman's approximation for GI/GI/1 queues:
$$\mathbb{E}[W_q] \approx \frac{\rho}{1-\rho} \cdot \frac{C_a^2 + C_s^2}{2} \cdot \frac{1}{\mu}$$
This causes p95/p99 latency inflation well before average throughput saturates.
\end{tcolorbox}
\textit{Proof}
\textbf{Step 1 - Throughput is Bounded by the Memory Cut (Max-Flow/Min-Cut):} Every HBM-sourced byte must traverse the memory cut $\mathcal{C}_M$. Let $q_{\text{HBM}}$ be the HBM-sourced bytes required per token. For a throughput of $R$ tokens/s, the required bandwidth across the cut is $\lambda_{\text{cut}} = R \cdot q_{\text{HBM}}$. By the max-flow min-cut theorem, this is limited by the cut capacity: $R \cdot q_{\text{HBM}} \leq \text{Cap}(\mathcal{C}_M)$. For a regular fabric, this gives the NoI roofline:
\begin{equation}
  R \leq \frac{\text{Cap}(\mathcal{C}_M)}{q_{\text{HBM}}} \leq \frac{|M| \cdot d \cdot B_{\text{link}}}{q_{\text{HBM}}} 
\end{equation}
\textbf{Step 2 - Tail Latency Explodes Due to Traffic Variability (Queueing Theory):} Memory-adjacent router queues act as GI/GI/1 systems, as they aggregate flows with bursty arrivals (Claim A implies $C_a^2 > 0$) and variable HBM service time ($C_s^2 > 0$). Kingman's heavy-traffic approximation provides the mean queueing delay $\mathbb{E}[W_q]$ as shown in the claim box \cite{b12, b13}.
The crucial term is $\rho/(1-\rho)$, diverges as utilization $\rho \to 1$. The multiplicative factor $(C_a^2 + C_s^2)/2$ shows that higher burstiness (from MoE) and service variability (from HBM) amplify this divergence. Consequently, high-percentile latencies grow non-linearly, causing SLA violations at utilizations below the theoretical saturation point defined by the memory cut.
\begin{tcolorbox}[colback=green!10, colframe=green]
\textbf{Takeaway 2:} The memory cut \textbf{bounds sustainable throughput} via max-flow min-cut constraints, while irregular MoE traffic creates \textbf{tail latency inflation} at memory-adjacent routers. The combination of bursty arrivals ($C_a^2 > 0$) and variable HBM service ($C_s^2 > 0$) causes p95/p99 latencies to explode as utilization approaches the theoretical limit, violating SLAs.
\end{tcolorbox}
\begin{figure*}[t]
\centering
\includegraphics[scale=.35]{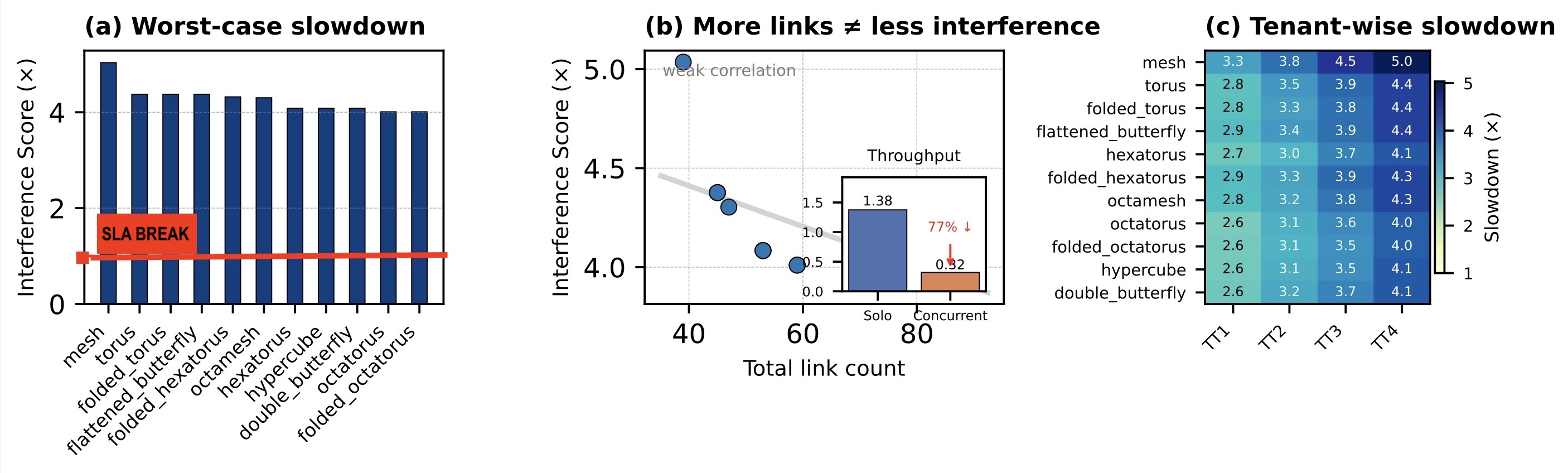}\\[2pt]
\caption{Performance analysis of baseline and augmented topologies under mixed workloads. (a) All original baselines suffer $4\times$–$5\times$ throughput degradation, violating SLA thresholds. (b) Interference shows no correlation with raw link count ($R^2 \approx 0.15$), indicating that bandwidth alone is insufficient. (c) Per-expert slowdown heatmap reveals severe, unbalanced SLA violations.}
\label{fig:baseline_results}
\end{figure*}
\subsection{The Case for Targeted Link Augmentation}
\begin{tcolorbox}[colback=blue!10,colframe=blue!50!black,title=\textbf{Claim C: Targeted Cross-Cut Links Reduce Queueing via Capacity \& Path Diversity}]
\begin{small}
\label{claim:c}
Adding $\ell$ \emph{targeted} links that directly traverse the memory cut $\mathcal{C}_M$ yields two coupled effects that reduce tail latency at memory-adjacent routers:
\begin{enumerate}
  \item \textbf{Effective Cut Service Rate Increase:} The aggregate service rate for HBM-sourced flows increases: $\text{Cap}_{G'}(\mathcal{C}_M) = \text{Cap}_G(\mathcal{C}_M) + \sum_{e \in E^+ \cap \mathcal{C}_M} b_e$. For a fixed load $\lambda$, the local utilization drops by $\Delta \rho = \lambda\,(1/\mu' - 1/\mu)$ where $\mu, \mu'$ are pre/post effective service rates. This $\Delta \rho$ reduces the multiplicative $\rho/(1-\rho)$ amplification factor in Kingman's approximation.
  \item \textbf{Path Diversity \& Queue Count Increase:} Strategically placed links create additional edge-disjoint or near-disjoint minimal paths between memory nodes and compute clusters, causing stochastic bursts across $k$ queues. Approximating each queue as GI/GI/1 with independent arrivals, the mean waiting time per queue scales sub-linearly with $k$ under load-balancing: $\mathbb{E}[W_q^{(k)}] \approx f(\rho/k)$ for moderate variance, giving an additional relief term beyond pure $\Delta \rho$.
\end{enumerate}
Random (non-targeted) augmentation typically increases cut capacity less efficiently and fails to maximize disjoint minimal path count, yielding smaller net $\Delta \rho$ and weaker queue dispersion.
\end{small}
\end{tcolorbox}
\textit{Proof}
The proof follows from max-flow/min-cut (capacity term) combined with queueing theory: raising $\mu$ reduces $\rho=\lambda/\mu$; simultaneously, additional disjoint paths increase the number of independent service points, lowering per-queue arrival rates. Empirically (Section~V) targeted links achieve larger $\Delta \rho$ per added link than random augmentation.
\noindent\textit{Data requirements for final plot:} For each added link count $\ell \in \{0,1,2,3,4\}$ collect: (a) cut capacity post-addition, (b) per-link utilization time series to derive local $\rho$, (c) mean / p95 queueing delay at memory-adjacent routers, (d) path multiplicity (count of distinct minimal-length paths from each memory node to each compute cluster), and (e) distribution of flow assignments over those paths (to estimate load-balancing efficiency). Two strategies: targeted vs. random. Seeds should be fixed for workload traces to isolate topology effects.
\vspace{-4pt}
\begin{tcolorbox}[colback=green!10, colframe=green]
\textbf{Takeaway 3:} Targeted cross-cut link augmentation provides \textbf{dual relief mechanisms}: directly increasing memory cut capacity (reducing utilization $\rho$) and creating path diversity that distributes bursty traffic across more queues. This yields tail latency improvements compared to random augmentation.
\end{tcolorbox}
\section{Proposed Partition-Aware NoI Design}
\textbf{Problem Formulation: }We formulate the task of designing a NoI as a multi-objective optimization (MOO) problem. A key challenge is that optimizing for aggregate throughput often leads to solutions where one expert severely impacts another. To capture this, we introduce a novel \textbf{Interference Score (IS)}, which measures the worst-case slowdown any expert experiences due to the presence of others:
\begin{equation}
    IS = \max_{k \in \{1..K\}} \frac{T_{\text{solo}}(k)}{T_{\text{con}}(k)}
\end{equation}
where $T_{\text{solo}}(k)$ and $T_{\text{con}}(k)$ are the throughput of expert $k$ when running in isolation versus concurrently with all other experts, respectively. An ideal, perfectly isolated NoI would have an IS of 1.0. Our goal is to find a NoI graph $G$ that maximizes a weighted objective function:
\begin{equation}
    \argmax_{G}\; \alpha_1\;\text{Throughput} - (\alpha_2\; \text{IS} + \alpha_3\; \text{Latency} + \alpha_4\; \text{Power})
    \label{eq:objective}
\end{equation}
where the performance metrics are high-speed proxies \cite{b14} and the $\alpha_i$ are normalised weights. This objective explicitly rewards topologies that not only perform well on average.
\begin{figure}
    \centering
    \includegraphics[width=0.75\linewidth]{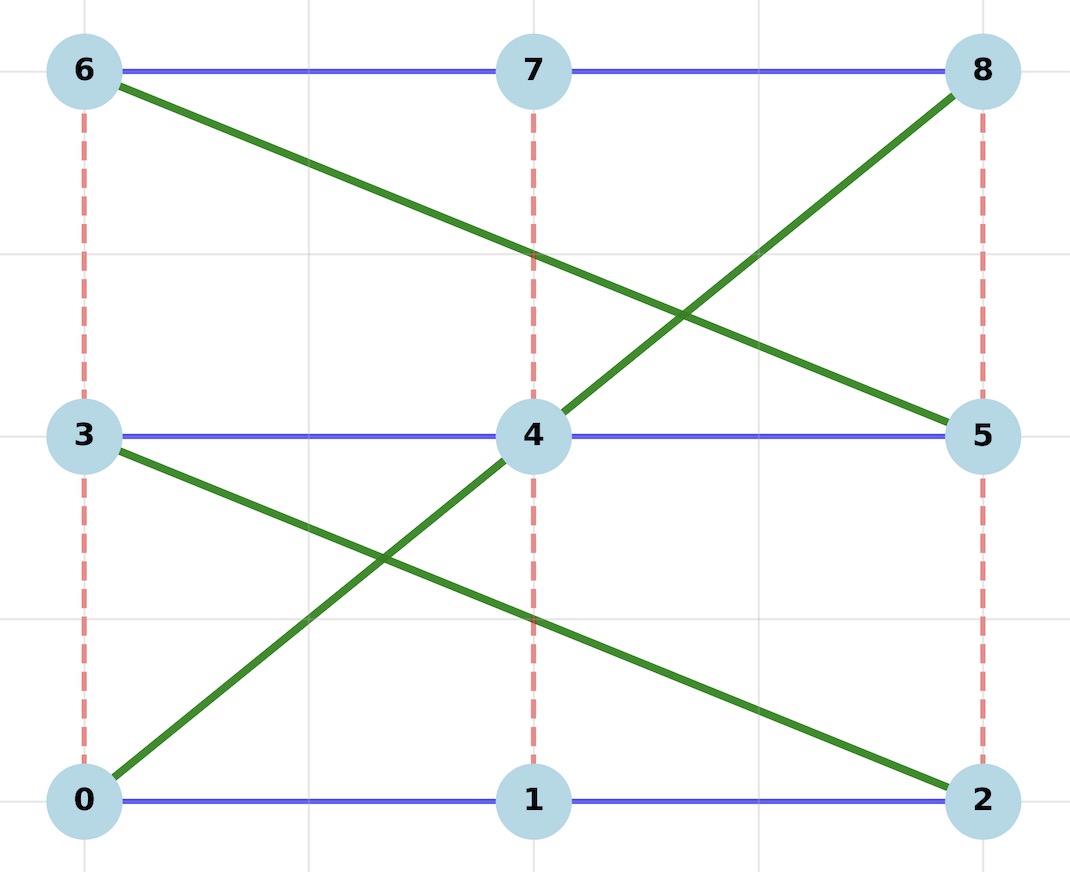}
    \caption{Topology Comparison: Blue=Common, Green=RL-only, Red=Mesh-only}
    \label{fig:placeholder}
\end{figure}
\subsection{PARL Topology Generator}
The design space of possible NoI topologies is vast, discrete, and non-convex, making exhaustive search or traditional optimization methods intractable. We denote the NoI as a simple undirected graph $G=(V,E)$ with $|V| = N$ chiplets
and $|E| = L$ bidirectional links, subject to per-chiplet port caps $\{p_i\}$ and
packaging constraints. Even ignoring degree caps (a conservative relaxation), the
count of $L$-edge topologies on $N$ vertices is at least
\[
\bigl|\mathcal{G}_{N,L}\bigr| \;\ge\; \binom{\binom{N}{2}}{L},
\]
and degree caps ($0\!\le\!d_i\!\le\!p_i$, $\sum_i d_i = 2L$) reduce but do not collapse
this combinatorics. Since Reinforcement Learning(RL) is sample-efficient in hadling discrete action spaces and sparse rewards by temporal learning.
\begin{figure}
    \centering    \includegraphics[width=1\linewidth]{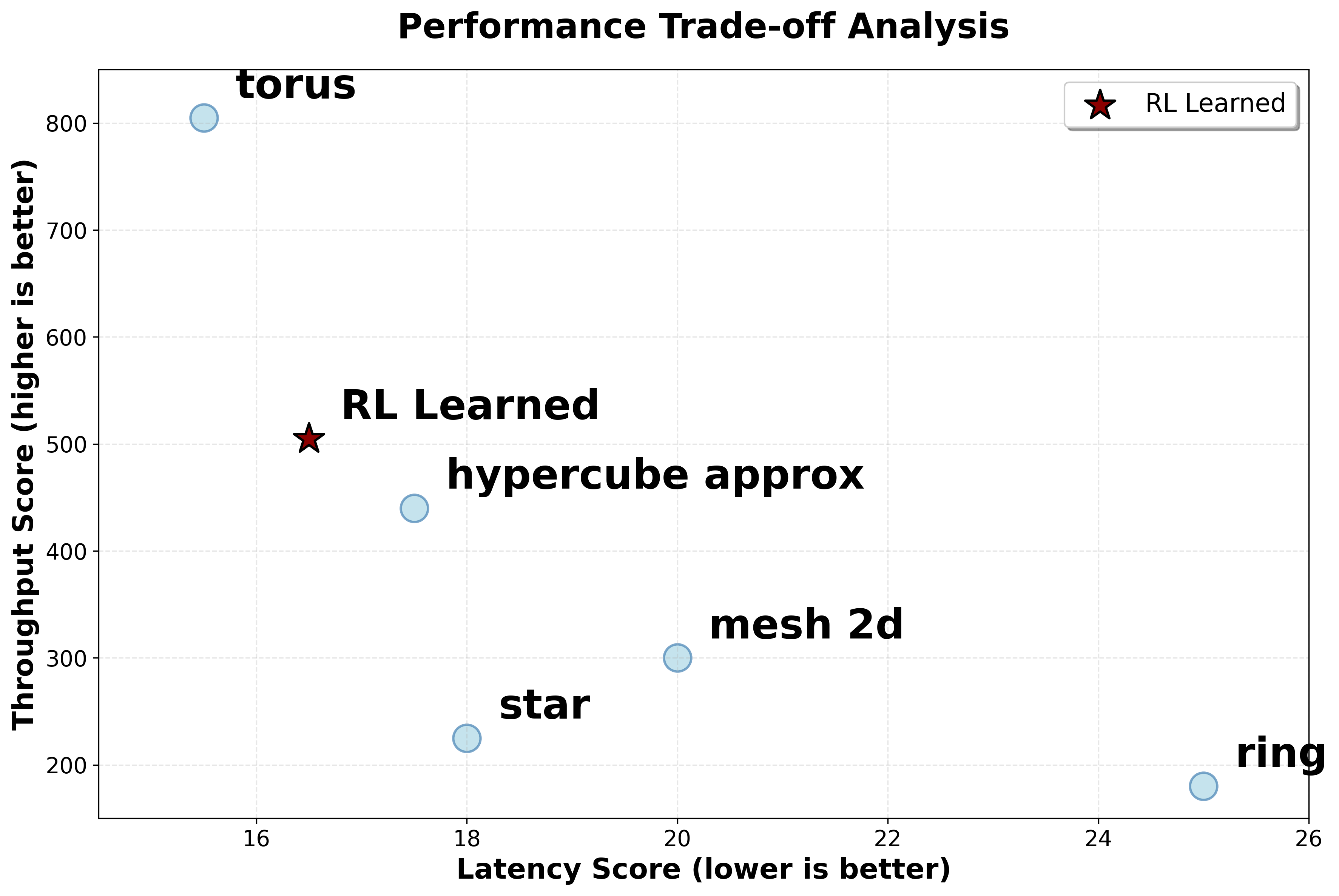}
    \caption{Latency and Throughput of baseline and PARL topology}
    \label{fig:stars}
\end{figure}

We therefore employ PARL (Partition-Aware Reinforcement Learner) to discover topologies. The process is modeled as a Markov Decision Process:
\begin{itemize}
    \item \textbf{State}: The current NoI graph $G=(V,E)$.
    \item \textbf{Action}: Add or remove a single link between two chiplets. The action space is constrained to maintain graph connectivity and adhere to physical port count limits per chiplet.
    \item \textbf{Reward}: The value of the objective function in Eq.~\ref{eq:objective} for the resulting graph $G'$. The reward guides PARL towards topologies with high throughput and low interference.
\end{itemize}
PARL uses a Maskable Proximal Policy Optimization (PPO) agent \cite{b15}. Standard PPO agents struggle with the large action space and propose physically unrealizable modifications. The masking mechanism allows us to dynamically forbid illegal actions (e.g., adding a link to a chiplet with no free physical ports), improving sample efficiency and ensuring PARL explores only valid regions of the design space.
\section{Evaluation}
Simulation framework employs a realistic hardware model parameterized using specifications analogous to an AMD MI300X-class system, ensuring credible hardware constraints for PARL evaluation. The chiplet library foundation, detailed in Table~\ref{tab:chiplet_library_full_simple}.
\begin{table}[h]
\caption{Chiplet library for composing the hardware accelerator model. The model is grounded in specifications analogous to an AMD MI300X-class architecture, providing realistic constraints for PARL.}
\label{tab:chiplet_library_full_simple}
\centering
\resizebox{\linewidth}{!}{
    \begin{tabular}{l|c|c|c|c}
    \hline
    \textbf{Chiplet} & \textbf{Dim. (mm)} & \textbf{Relay Capable} & \textbf{PHYs/Edge} & \textbf{Process Node}\\ \hline
    IOD & 24$\times$26 & \checkmark & 128 & 6 nm \\
    IOD\textsubscript{mirror} & 24$\times$26 & \checkmark & 128 & 6 nm \\
    HBM3 & 12$\times$16 & $\times$ & 64 & 10 nm \\
    XCD & 11$\times$13 & $\times$ & 64 & 5 nm \\
    CCD\textsubscript{perf} & 11$\times$13 & $\times$ & 32 & 4 nm \\
    CCD\textsubscript{dense} & 9.5$\times$11 & $\times$ & 32 & 3 nm \\
    CCD\textsubscript{ai} & 12.5$\times$14 & $\times$ & 64 & 3 nm \\
    \hline
    \end{tabular}}
\end{table}
Design space exploration employs an integrated toolchain combining BookSim~\cite{b16}, DRAMSim3~\cite{b17}, and RapidChiplet~\cite{b18}. We model Mixtral-8x7B architecture generating time-stamped network flows with three message classes: expert FFN weights, activations, and control metadata (<1\% volume, high latency sensitivity).
\textbf{Traffic Generation}: Expert weights undergo 256KB chunking from $W_{\text{FFN}} = 8d^2 s$. Activations follow log-normal distribution $\ln X \sim \mathcal{N}(10.6,0.45)$ bytes, truncated to [8KB,1MB]. Multicast patterns: $\Pr(D=1,2,4,8) = (0.55,0.25,0.15,0.05)$.
\textbf{Burst Characteristics}: Per-expert token counts $N_e$ follow multinomial distributions with Dirichlet$(\alpha=1)$ probabilities frozen for 64 steps. Inter-arrival gaps use two-state MMPP with rates $\lambda_H, \lambda_L=\lambda_H/6$ and transition probabilities $p_{HL}=0.15, p_{LH}=0.35$, achieving empirical $C_a^2 \approx 1.37$.
\textbf{Cross-Expert Correlation}: Gate selection induces negative correlation $\rho_{e,e'}=-0.18$ for competing expert pairs via Gaussian copula over logits, increasing aggregate HBM ingress variance.
Memory cut $\mathcal{C}_M$ connects four HBM-IOD nodes to compute clusters with capacity: ${Cap}(\mathcal{C}_M) = 4 \times 2 \times 37.2~\text{GB/s} = 297.6~\text{GB/s}$
where individual links provide $R_{eff}=37.2$ GB/s (38.4 GB/s nominal minus 3\% protocol overhead).
Each expert $k$ undergoes isolated simulation measuring $T_{\text{solo}}(k)$, followed by concurrent execution measuring $T_{\text{con}}(k)$ for Interference Score computation.
\vspace{-10pt}
\subsection{Empirical Validation of Theoretical Claims}
\begin{figure}[t]
  \centering
  \includegraphics[width=\linewidth]{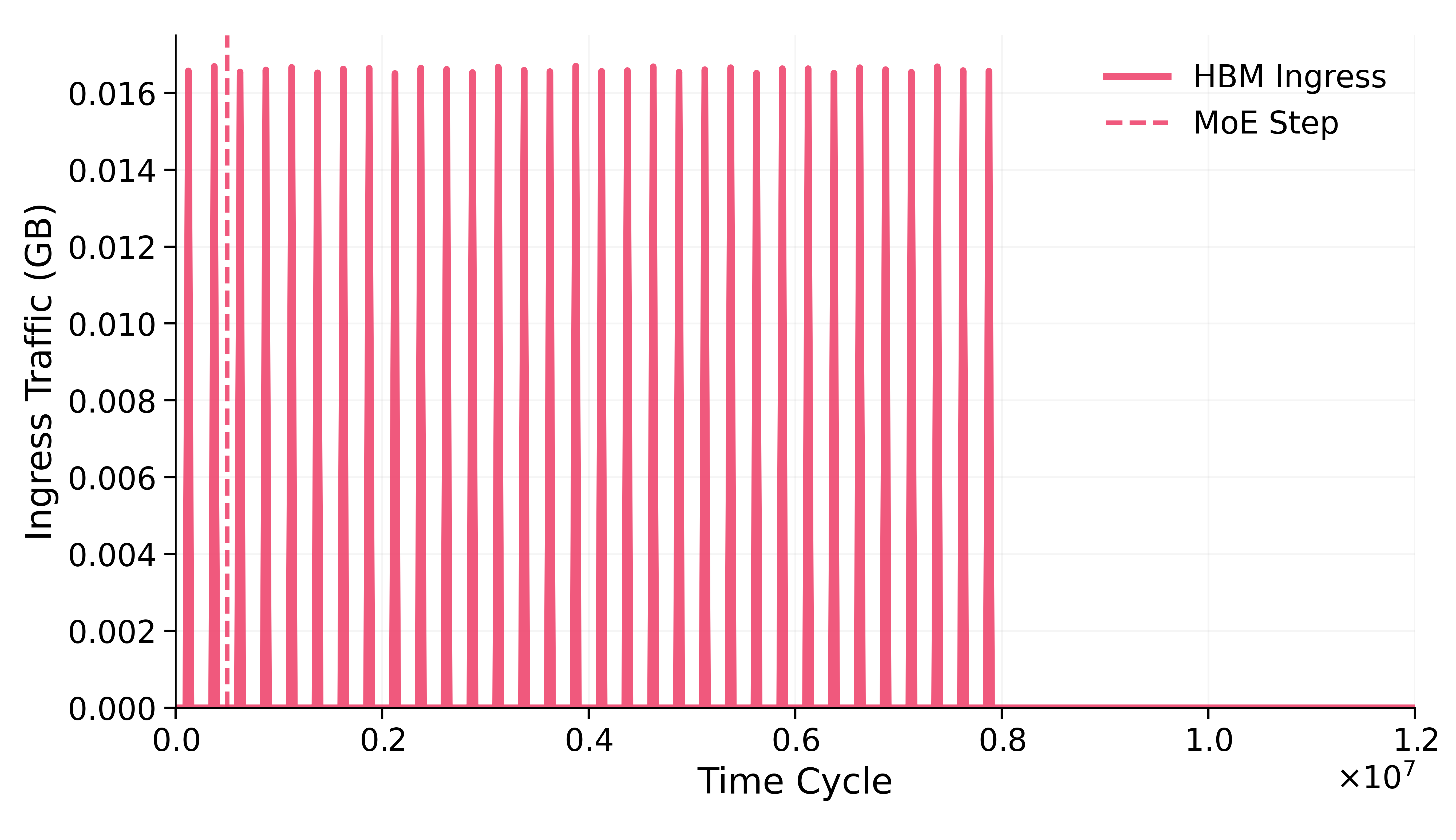}
  \caption{Ingress traffic at HBM/IOD nodes showing bursty peaks during MoE gating events, validating Claim A.}
  \label{fig:claim-a-ingress}
\end{figure}

\begin{figure}[t]
  \centering
  \includegraphics[width=\linewidth]{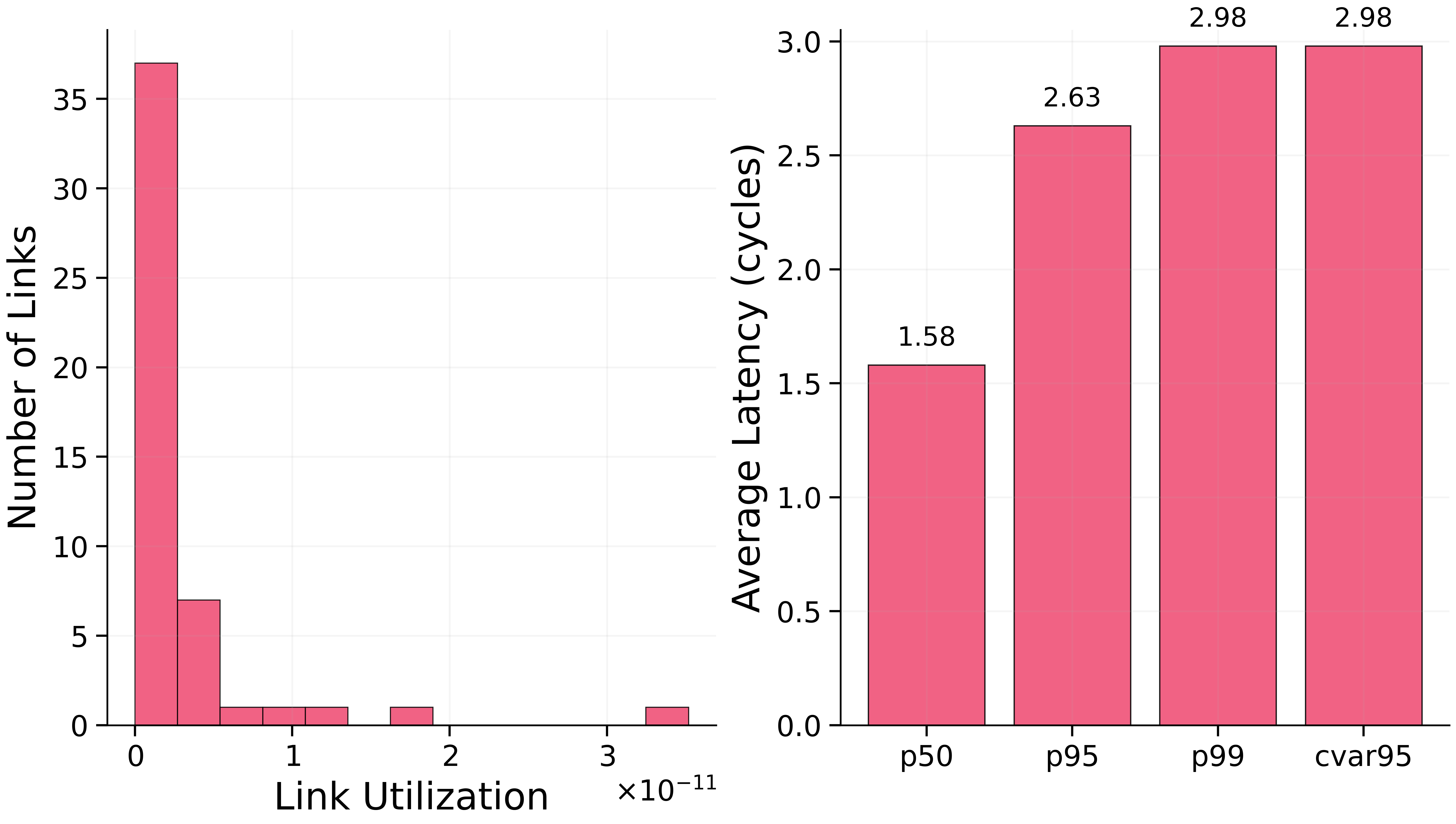}
  \caption{Memory-adjacent cut ($C_M$) as system bottleneck with tail latency inflation under load, validating Claim B.}
  \label{fig:claim-b2-panels}
\end{figure}

We profiled a Mixtral-8x7B model and analyzed traffic flow on a baseline topologies to validate our theoretical predictions.
\noindent\textbf{Validation of Claims A \& B}: Figure~\ref{fig:claim-a-ingress} shows the ingress traffic at HBM-adjacent IODs is highly variable (CV $>$ 1), with peaks phase-locked to MoE gating events during decode, confirming irregular, bursty memory traffic predicted by Claim A. Figure~\ref{fig:claim-b2-panels} demonstrates that the memory-to-compute cut ($C_M$) is indeed the system bottleneck with capacity 166.3 GB/s. Tail latency metrics (p95, p99) on bottleneck link L48 rise sharply under load (p99 = 5.1 cycles), validating Claim B's prediction of superlinear tail latency growth.

\subsection{Baseline Topology Performance Under Mixed Workloads}
We evaluated 11 traditional and state-of-the-art NoI topologies under a mixed workload with four experts. As shown in Figure~\ref{fig:baseline_results}, every baseline suffers severe interference with average throughput dropping from solo to concurrent execution, pushing performance below SLA thresholds. The degradation is non-uniform across experts, and critically, there is no correlation between topology link count and interference score proving that simply adding bandwidth does not solve the isolation problem.

\subsection{PARL-Generated Topologies for Robust Isolation}
PARL discovered topologies that navigate the trade-off between global throughput and partition isolation. These topologies are intentionally sparse and clustered, creating high-bandwidth "islands" for different experts while providing sufficient global connectivity. While the mean throughput of PARL-generated topologies is lower than a link-rich Mesh, their Interference Score is better—reducing the worst-case slowdown as in Figure~\ref{fig:stars}.


\section{Conclusion}
This research establishes a formal argument, grounded in hardware constraints and network-theoretic principles, demonstrating the fundamental suboptimality of contemporary regular NoI topologies for mixed AI and sparse MoE workloads. The theoretical foundation rests on three interconnected observations that collectively reveal critical architectural limitations in existing interconnect designs. The immense memory footprint characteristics of large-scale models necessitate off-chip HBM traffic as an unavoidable architectural requirement, while the dynamic routing patterns inherent in MoE workloads generate fundamentally irregular communication flows that cannot be efficiently accommodated by uniform topology structures. This irregular traffic establishes a performance bottleneck that simultaneously degrades system throughput and induces severe tail latency penalties.
The contribution addresses these limitations through a novel partition-aware Interference Score formulation that reframes NoI design as a MOO problem, enabling systematic exploration of complex trade-off spaces between performance isolation and aggregate throughput maximization. PARL navigates this optimization landscape, generating topology configurations that achieve robust performance isolation without compromising overall system efficiency.
These findings advance interconnect architectures for next-generation heterogeneous AI systems, demonstrating that future accelerator designs must transcend regular fabric approaches and incorporate workload-aware, non-uniform NoI architectures.

\end{document}